%% file: 0-main.tex
\documentclass[letterpaper]{article} 
\usepackage{aaai2026}  
\usepackage{times}  
\usepackage{helvet}  
\usepackage{courier}  
\usepackage[hyphens]{url}  
\usepackage{graphicx} 
\usepackage{natbib}  
\usepackage{caption} 
\usepackage{algorithm}
\usepackage{algorithmic}

\usepackage{newfloat}
\usepackage{listings}
\DeclareCaptionStyle{ruled}{labelfont=normalfont,labelsep=colon,strut=off} 
\floatstyle{ruled}
\newfloat{listing}{tb}{lst}{}
\floatname{listing}{Listing}
\usepackage[table]{xcolor}
\usepackage{tabularx}
\usepackage{booktabs}
\usepackage{subcaption}
\usepackage{amsmath}
\usepackage{amssymb}
\usepackage{tikz}
\usepackage{multirow}

\newcolumntype{Y}{>{\raggedright\arraybackslash}X} 

\defcitealias{trujillo2025dsa}{Trujillo et al. 2025}
\defcitealias{dsa2023tdb}{EC 2023}

\definecolor{inadequate}{HTML}{AA0000} 
\definecolor{semiadequate}{HTML}{CCAAAA} 
\definecolor{adequate}{HTML}{EAEAEA} 

\title{Disarranged Harmonization of Transparency Reporting\\by Social Media Platforms Under the Digital Services Act\footnote{\textcolor{red}{Accepted to ICWSM 2027.}}}
\author{
    Amaury Trujillo\textsuperscript{\rm 1}
    Benedetta Tessa\textsuperscript{\rm 2, \rm 1},
    Stefano Cresci\textsuperscript{\rm 1}
}
\affiliations{
    \textsuperscript{\rm 1}IIT-CNR, Pisa, Italy\\
    \textsuperscript{\rm 2}University of Pisa, Italy\\
    amaury.trujillo@iit.cnr.it, benedetta.tessa@phd.unipi.it, stefano.cresci@iit.cnr.it
}

\begin{document}

\maketitle

\begin{abstract}
The European Commission recently introduced new regulation to harmonize transparency reporting of large online platforms under the Digital Services Act (DSA). Here, we present the first systematic evaluation of transparency reporting data quality after this normative change, for the eight largest social media platforms in the European Union. In detail, we run a set of large-scale quantitative analyses on key reporting dimensions, followed by a structured comparative assessment across platforms and reporting mechanisms. Among our findings is that: \textit{(i)} the analyzed platforms had varying degrees of compliance and data quality, but all exhibited issues on data formatting, timeliness, consistency, and completeness; \textit{(ii)} some platforms employed differing reporting procedures across mechanisms, which caused them to submit contrasting information; \textit{(iii)} despite the harmonization, a number of issues still prevent interoperability between reporting mechanisms; and \textit{(iv)} many of the previously identified issues with transparency reporting are still unresolved. We conclude by discussing implications for transparency auditing and proposing key targeted improvements to strengthen the reliability and interoperability of DSA transparency reporting.
\end{abstract}

\input{1-introduction}
\input{2-related-work}
\input{3-approach}
\input{4-analyses}
\input{5-evaluation}
\input{6-discussion}

\section{Acknowledgents}
This work is partly supported by the European Union with the ERC project DEDUCE under grant \#101113826.

\bibliography{aaai2026}

\input{a-appendix}

\end{document}

%% file: 1-introduction.tex
\section{Introduction}


The Digital Services Act (DSA) is arguably the most ambitious European Union (EU) regulation to institutionalize the transparency and accountability of online platforms. Its impact extends beyond the European regulatory context, however, bearing global governance implications in platform regulation via the so-called \emph{Brussels effect}~\cite{elkin2025regulating}. Platform transparency under the DSA is thus an operational prerequisite for independent auditing, regulatory oversight, and public accountability about content moderation practices, with the aim of a safe and trusted online environment.
To achieve its goal, the DSA introduced two complementary reporting mechanisms: the Transparency Database (\texttt{TDB}) and Transparency Reports (\texttt{TRs}).
On one hand, the \texttt{TDB}\footnote{\url{https://transparency.dsa.ec.europa.eu/}} is a centralized public repository of \emph{statements of reasons} (SoRs) submitted by platforms for each individual moderation action affecting users. These records provide itemized metadata with the grounds, targets, and procedural characteristics of moderation actions~\cite{trujillo2025dsa,kaushal2024automated,papaevangelou2024content}. On the other hand, \texttt{TRs} are documents published periodically through which platforms disclose aggregated information on moderation practices, such as automated systems, notice handling, and risk mitigation measures~\cite{urman2023transparent}. Together, the \texttt{TDB} and \texttt{TRs} form a dual transparency framework intended to facilitate public scrutiny, regulatory oversight, and cross-platform comparability in content moderation, by respectively offering low and high levels of detail.
In addition, very large online platforms (VLOPs) ---those with more than 45M users in the EU--- must provide additional details in their reporting, in view of their greater impact.

Despite the unprecedented scope and public accessibility, early studies of these mechanisms identified a range of limitations affecting the interpretability, comparability, and reliability of the data reported by VLOPs. Concerning the \texttt{TDB}, research surfaced heterogeneous reporting practices across platforms, extensive use of ambiguous or generic categories, missing contextual information, and inconsistencies between structured fields and free-text descriptions~\cite{trujillo2025dsa,kaushal2024automated}. Similarly, studies of the \texttt{TRs} highlighted strong differences in reporting formats, aggregation criteria, and methodological explanations, which hindered systematic cross-platform analyses~\cite{urman2023transparent,dhar2023impact}. In response to these limitations, the European Commission (EC) introduced a major harmonization effort for the \texttt{TRs} effective 1 July 2025.\footnote{\url{https://digital-strategy.ec.europa.eu/en/library/implementing-regulation-laying-down-templates-concerning-transparency-reporting-obligations}} The first harmonized reports thus cover the second half of the year (2025H2). The initiative standardized the structure, formatting, terminology, and content requirements of the \texttt{TRs} through official templates and binding instructions in line with broader efforts aimed at improving comparability, machine readability, and interoperability of reporting practices across platforms~\cite{bommasani2024foundation, reid2024transparency}. At the same time, the \texttt{TDB} was also slightly changed to align its taxonomy and structure with the new harmonized \texttt{TRs}.\footnote{\url{https://transparency.dsa.ec.europa.eu/page/api-documentation?lang=en#changelog}} These changes represent a considerable attempt to operationalize a standardized transparency infrastructure under the DSA. However, because these mechanisms have been only recently introduced, there is currently no empirical evidence on how well the platforms complied with the new reporting specifications, whether harmonization effectively improved cross-platform comparability, or whether the long-standing inconsistencies previously identified in both transparency mechanisms have been mitigated.

For these reasons, we examine the first harmonized \texttt{TRs} submitted by the largest EU social media platforms for 2025H2, and systematically compare them against the corresponding data available in the \texttt{TDB}. Following recent scholarly proposals~\cite{tessa2025improving} and the very rationale underlying the harmonization effort, we conduct a quantitative analysis of the \texttt{TRs} to assess their formal compliance with the new specifications and their practical interoperability with the database. While prior work already performed manual cross-checking between the reports and the database~\cite{trujillo2025dsa}, the introduction of common and machine-readable formats creates, for the first time, the conditions for comprehensive programmatic auditing. Our analysis therefore evaluates the extent to which such operationalization is currently feasible and identifies the current limitations that hinder systematic auditing. Thus, beyond revisiting previously identified issues, our main contribution is to evaluate whether the EC's harmonization intervention has translated formal standardization into consistent, interoperable, and systematically auditable transparency reporting in practice. Our work also differs from previous evaluations of transparency reporting~\cite{urman2023transparent} since these evaluated transparency against external and non-regulatory frameworks~\cite{principles2021santa}, while here we assess it directly against the regulatory specifications established by the EC itself. Finally, our analysis also reevaluates the current state of the database nearly two years after its launch, assessing whether the inconsistencies and reporting issues identified in early studies have been mitigated~\cite{kaushal2024automated,trujillo2025dsa,shahi2025year}. Accordingly, this work addresses the following guiding research questions:
\begin{itemize}
    \item \textbf{RQ1:} To what extent do the harmonized Transparency Reports  and Transparency Database records comply with the European Commission’s instructions?
    \item \textbf{RQ2:} To what extent are harmonized Transparency Reports data consistent with the corresponding Transparency Database data?
\end{itemize}
Beyond their methodological relevance, these questions also have broader implications for the governance and enforceability of the DSA itself and, through the Brussels effect, for its potential influence on transparency standards beyond the EU. Several formal proceedings initiated by the EC against major platforms already involve alleged failures concerning regulatory and transparency compliance. In this context, assessing the consistency, interoperability, and reliability of DSA transparency data is not only necessary to evaluate platform reporting practices, but also to understand the practical capacity of the EC and the DSA framework to support meaningful enforcement at scale.

%% file: 2-related-work.tex
\section{Related Work}


\subsection{Transparency Database}
The \texttt{TDB} sparked great interest among scholars around its release, but ---despite its potential--- several studies highlighted reliability issues in its data. For example, platforms frequently resort to broad or generic categories when describing the grounds for moderation decisions, and predominantly refer to violations of their terms of service without specifying the exact provision
~(\citealt{kaushal2024automated}; \citealt{papaevangelou2024content}; \citealt{drolsbach2024content}; \citetalias{trujillo2025dsa}). There is also a limited use of optional attributes that could contextualize moderation decisions~\citepalias{trujillo2025dsa}. These issues persisted both in the early stages of the database and one year after its launch~\cite{shahi2025year,tessa2026transparency}.
Some challenges also arise from the database structure and reporting practices themselves, including the lack of explicit schema representations for certain interventions, such as actions on accounts~(\citealt{groesch2025big}; \citetalias{trujillo2025dsa}).
In addition, various inconsistencies were observed between the \texttt{TDB} and other reporting mechanisms. For example, X, Facebook, and Instagram disclosed great use of solely automated moderation in their \texttt{TRs}, while their corresponding \texttt{TDB} actions were not fully automated~(\citetalias{trujillo2025dsa}; \citealt{kaushal2024automated}). To address these issues,~\citet{tessa2025improving} proposed two complementary verification processes to assess the consistency and reliability of DSA transparency data through systematic cross-checking between \texttt{TRs}, \texttt{TDB} records, and internal platform data. Our work follows this direction by operationalizing cross-checking of the newly-harmonized \texttt{TRs}.

Beyond assessing data quality and consistency, recent work has also explored methods to improve the database's usability and interpretability. For example, multi-agent systems based on large language models (LLMs) have been used to link SoRs to relevant sections of platforms’ terms of service~\cite{aspromonte-etal-2024-llms}. Similarly,~\citet{esser-spanakis-2025-linking} explored methods for linking SoRs to the most relevant clauses in platform policy documents. Using a dataset of TikTok SoRs, they evaluated several retrieval and language models for mapping moderation decisions to policy text, demonstrating how such approaches can improve the interpretability of moderation actions and support the analysis of enforcement practices.

\subsection{Transparency Reports}
\texttt{TRs} have also been scrutinized for their heterogeneous reporting practices across platforms. \citet{urman2023transparent} showed that these reports often provide aggregated moderation statistics with limited methodological detail, while the scope and granularity of disclosed information vary substantially across platforms. By evaluating \texttt{TRs} against the Santa Clara Principles on transparency and accountability in content moderation~\cite{principles2021santa}, they further highlighted significant inconsistencies in reporting practices, including uneven disclosure of government requests and moderation activities, ultimately revealing a broader lack of standardization. These inconsistencies make it difficult to assess whether platforms provide an accurate representation of their moderation practices, particularly considering that moderation effectiveness may depend on operational factors, such as the timing of interventions, that are often absent from aggregated reports~\cite{trujillo2025dsa,truong2025delayed}. Independent oversight bodies have further noted that certain interactions, particularly those involving law enforcement, may be insufficiently represented in these reports~\cite{van2025article}. Importantly, these works analyzed \texttt{TRs} published before the EC introduced the harmonized reporting framework and its associated standardization requirements.

\subsection{Broader Challenges in DSA Oversight}
Beyond the issues with the \texttt{TDB} and \texttt{TRs}, prior research also highlighted several challenges affecting the broader reporting and oversight ecosystem, emphasizing that the evaluation of DSA mechanisms remains methodologically difficult due to persistent gaps between regulatory objectives and empirically observable platform behavior~\cite{nannini2025beyond}. These limitations are also reflected at the level of specific enforcement mechanisms. For instance, the ``Notice and Action'' process (Article 16 of the DSA) may place a substantial burden on users, who are often required to classify potentially illegal content despite lacking formal legal expertise~\cite{sekwenz2025unfair}. Similar constraints affect the work of Trusted Flaggers (Article 22), whose limited resources can complicate the balance between effective reporting, institutional legitimacy, and the risk of over-removal~\cite{van2025article,sekwenz2026there}. At the same time, researcher access provisions (Article 40) have faced substantial implementation barriers~\cite{jaursch2024dsa}, despite their potential utility for independent audits and transparency assessments~\cite{tessa2025improving}, with persistent information asymmetries between platforms and academia limiting meaningful external scrutiny~\cite{goanta2025great}. Our work contributes to this broader effort by assessing whether the recent harmonization initiative effectively advanced transparency, interoperability, and independent oversight under the DSA.

%% file: 3-approach.tex
\section{Data}
\label{sec:data}
We focus on the newly-harmonized transparency reports for the second semester of 2025 (2025H2) published by the eight social media platforms originally designated as very large online platforms (VLOPs)---namely, Facebook, Instagram, LinkedIn, Pinterest, Snapchat, TikTok, X, and YouTube. We first collected all 630.27M SoRs sent by the above eight platforms to the \texttt{TDB} during 2025H2, as reported in Appendix Table~\ref{tab:sor_per_amar}. These SoRs are detailed and itemized, but only contain metadata about moderation actions. As such, they do not include personally identifiable information, nor do they allow direct association with specific pieces of content or users. Next, we collected all the standardized files with aggregated data from the \texttt{TRs} covering 2025H2 from the respective platform repositories, as officially submitted to the EC.\footnote{\url{https://digital-strategy.ec.europa.eu/en/policies/dsa-brings-transparency}} The transparency data and documentation we used are publicly and freely available under a Creative Commons Attribution 4.0 International license~\citepalias{dsa2023tdb}, easing the study's reproducibility and extensions.
The main data and analysis code required to reproduce the findings of this study are publicly available\footnote{\url{https://doi.org/10.6084/m9.figshare.33794503}} at FigShare.

\begin{figure*}[t]
\centering
\includegraphics[width=0.9\textwidth]{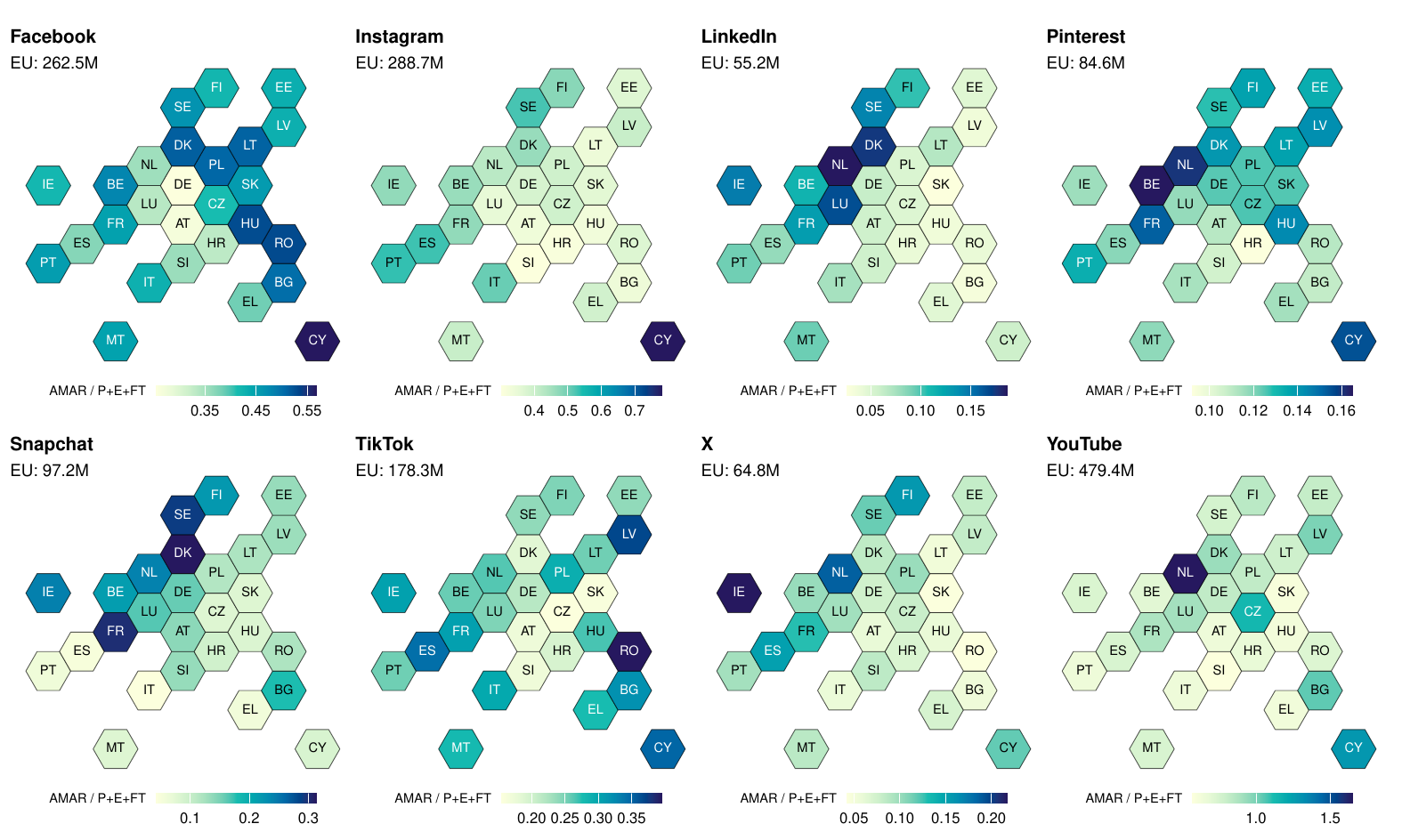} 
\caption{Platform-wise, country-level number of average monthly active recipients (AMAR), for the second half of 2025. AMAR are normalized by national population (P) plus active enterprises (E) and average monthly foreign tourists (FT).}
\label{fig:amar_by_platform_and_country}
\end{figure*}

\section{Methodology}
\label{sec:method}
Our methodological framework consists of two complementary layers: a set of quantitative analyses aimed at assessing key dimensions of post-harmonization transparency reporting, followed by a structured comparative assessment that critically synthesizes these results to characterize the degree of compliance, transparency, and interoperability across platforms and reporting mechanisms.

The quantitative analyses examine five key dimensions of platform reporting: \textit{(i)} number of active recipients (i.e., users), \textit{(ii)} timeliness of submitted SoRs, \textit{(iii)} use of automated means for content moderation, \textit{(iv)} moderation actions taken on the platforms’ own initiative, and \textit{(v)} actions resulting from notice mechanisms. These aspects correspond to central DSA transparency obligations and capture core operational components of platform moderation and governance~\cite{trujillo2025dsa,kaushal2024automated}. For each dimension, we first outline the relevant \emph{regulation} and normative references defining the corresponding reporting obligations, then present our quantitative \emph{analysis}, and finally discuss the resulting main \emph{findings}. Anchoring the analyses in the relevant regulatory provisions allows us to operationalize the corresponding DSA reporting obligations, systematically assess the extent of platform compliance based on the actual EC obligations, and clearly identify inconsistencies, omissions, and reporting limitations across transparency mechanisms. While additional analyses beyond the five considered here were also possible---some of which are reported in the Appendix---we focused on these dimensions because of their centrality and because they effectively illustrate broader patterns and limitations in the current transparency reporting ecosystem.

Next, we build on the results of the quantitative analyses to carry out a structured comparative assessment of the reporting quality of each platform across both transparency mechanisms. Specifically, we assigned scores along six evaluation dimensions---three for the \texttt{TDB} and three for \texttt{TRs}---covering aspects such as formatting, timeliness, consistency, and completeness. Finally, we discuss the main factors underlying the assigned scores for each platform. Therefore, rather than directly interpreting platform moderation practices, this assessment focuses on the extent to which the reporting itself complies with and operationalizes the harmonized specifications and requirements introduced by the EC, both within and across transparency mechanisms. Indeed, evaluating the consistency, completeness, and interoperability of transparency data is a necessary prerequisite for reliably analyzing the moderation practices these data are meant to describe. This critical synthesis enables a comparative assessment of reporting quality, harmonization adherence, and interoperability across both platforms and transparency mechanisms. 

In the following sections, for simplicity, we use \emph{statement of reasons} (SoR) to refer to data in the Transparency Database (\texttt{TDB}) and \emph{self-report} to data in files of the Transparency Reports (\texttt{TRs}). In addition, any mention of an Article or Recital refers to those in the DSA.

%% file: 4-analyses.tex
\section{Quantitative Analyses}
\label{sec:analyses}

\subsection{Active Recipients}
Average monthly active recipients (AMAR) are used to determine VLOP designation and several downstream regulatory obligations. As such, reporting inaccuracies may have broad implications for platform oversight and enforcement.

\paragraph{Regulation}
Article 3(p) defines an \emph{active recipient of an online platform} as a ``recipient of the service [natural or legal person] that has engaged with an online platform by either requesting the online platform to host information or being exposed to information hosted by the online platform and disseminated through its online interface''. Article 33(1) designates as VLOPs those platforms with AMAR in the EU equal or higher to 45M. Article 42(3) mandates that VLOPs provide AMAR by each EU member state. Recital 77 indicates that the DSA does not require specific tracking of individuals online for AMAR, but it also states that recipients using different interfaces (e.g., websites or apps) or mechanisms (e.g., different URLs) should be counted only once, and that bots should be discounted.


\begin{figure*}[t]
\centering
\includegraphics[width=1\textwidth]{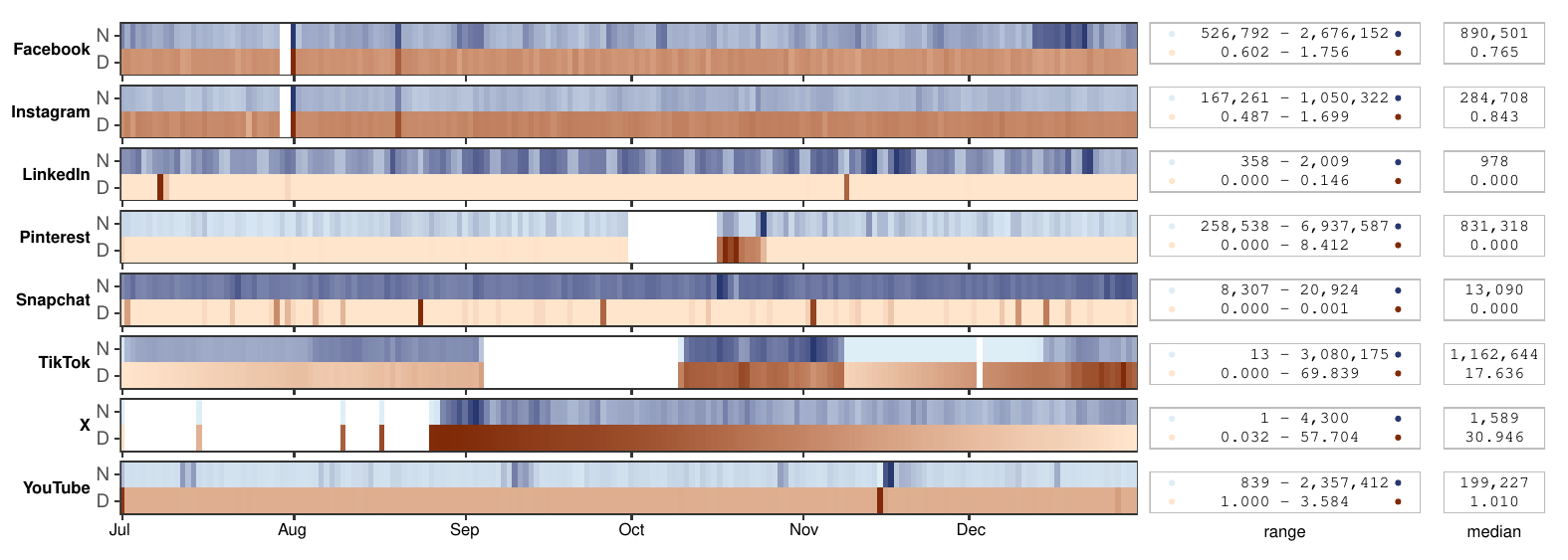} 
\caption{Platform-wise daily number of SoR submissions (N) and mean communication delay (D), during the second half of 2025. The right-hand panel reports the corresponding range and median values for each platform and metric. Delay is computed in days, from moderation decision to SoR submission. Blank spaces mark days in which platforms did not submit any SoR. Darker shades of color indicate higher numbers of submitted SoRs (N) or longer communication delays (D).}
\label{fig:daily_sor_count_and_delay}
\end{figure*}

\paragraph{Analysis}
We analyzed the country-level figures for AMAR disclosed by the platforms in their harmonized \texttt{TRs}. Concretely, we checked whether platforms computed AMAR consistently with the DSA obligations and whether the reported figures exhibit plausible scaling patterns with respect to country-level demographic and socioeconomic indicators. The cartograms in Figure~\ref{fig:amar_by_platform_and_country} depict platform-wise normalized AMAR per EU member state. The per-country normalization uses official Eurostat data\footnote{\url{https://ec.europa.eu/eurostat/databrowser/}} and accounts for 2025H2 national population, active enterprises, and average monthly foreign tourists, thereby covering natural persons, legal persons, and temporary visitors potentially qualifying as recipients under the DSA. This deliberately generous denominator provides a conservative upper-bound proxy rather than an estimate of actual recipients. As such, normalized AMAR values substantially above 1 would be unexpected. Figure~\ref{fig:amar_by_platform_and_country} shows that relative distributions of active recipients vary by platform and region. For example, LinkedIn and Snapchat are relatively more accessed from Northern and Western EU countries. More importantly, the figure reveals a few marked anomalies. Instagram, X, and YouTube exhibit countries with normalized AMAR values that are statistical outliers according to Tukey's interquartile-range criterion, which we use as a robust, distribution-free rule to flag unusually large observations without assuming them to be erroneous. For YouTube, the Netherlands exhibit a striking ratio of 1.65, corresponding to 47M AMAR per 18M residents, 8.5M enterprises, and 2M monthly foreign tourists. YouTube's parent company, Google, decries in its report the lack of precise EC guidance for calculating AMAR. However, it also states that AMAR should generally not exceed the sum of population and active enterprises, which contradicts its own country-level figures. The report further attributes possible overcounting to privacy-preserving limitations in cross-device tracking, despite the disclosed AMAR values referring to signed-in users.

\paragraph{Findings}
Platforms appear to operationalize AMAR inconsistently, with some reported figures difficult to reconcile with both DSA methodological indications and the platforms’ own reporting criteria. These results suggest that ambiguities in AMAR reporting persist despite harmonization.

\subsection{Timeliness of Statements of Reasons}
Timely submission of SoRs is required and essential for meaningful transparency and oversight under the DSA.

\paragraph{Regulation}
Article 24(5) states that platforms must submit their SoRs ``without undue delay'' to the \texttt{TDB}.

\paragraph{Analysis}
For each platform, Figure~\ref{fig:daily_sor_count_and_delay} depicts the time series of the daily number of submitted SoRs (top) together with the corresponding communication delay (bottom), measured as the number of days between moderation decisions and SoR submission to the \texttt{TDB}. Sustained interruptions in submissions or elevated communication delays may indicate operational difficulties, delayed reporting practices, or temporary failures in transparency compliance. Several platforms experienced substantial disruptions during 2025H2. Facebook and Instagram---both Meta platforms which appear to share the same SoR infrastructure---did not submit any SoR for two days at the end of July, but quickly recovered the delay on 1 August. LinkedIn continued to have the strong SoR weekly seasonality described by Trujillo et al. \citeyearpar{trujillo2025dsa} and practically had a zero-day communication delay. Pinterest lacked submissions for 16 days at the start of October, but recovered its delay the following week. Snapchat maintained a relatively stable rate of daily SoRS and zero-day delay. TikTok manifested a few interruptions or significant decreases throughout the six-month period, having 36 days with no submissions and high communication delays (median of 17.6 days). X is remarkable due to the virtual lack of submissions during the first 56 days, in which only 4 days had a handful of SoRs. At the end of July it started to recover, but it decreasingly spread the workload across the remaining days, resulting in the highest communication delays, with median delay of 30.9 days. YouTube had near-continuous daily submissions and generally exhibited a one-day communication delay. However, its submission volume showed substantial variability, including an unusually large two-day spike in mid-November that was not accompanied by a corresponding increase in communication delay.

\paragraph{Findings} 
The timeliness and continuity of SoR submissions varied substantially across platforms, with some exhibiting prolonged interruptions and communication delays in contrast with the DSA requirement to submit SoRs without undue delay.

\subsection{Use of Automated Means}
\label{sec:automated_means}
The use of automated means for content moderation is central to the DSA framework and directly affects the interpretability and accountability of platform moderation practices. Moreover, the reliable disclosure of automated moderation is also relevant under the EU AI Act,\footnote{\url{https://eur-lex.europa.eu/eli/reg/2024/1689/oj/eng}} which introduces additional governance and transparency obligations for AI systems deployed at scale.

\begin{figure*}[t]
\centering
\includegraphics[width=.8\textwidth]{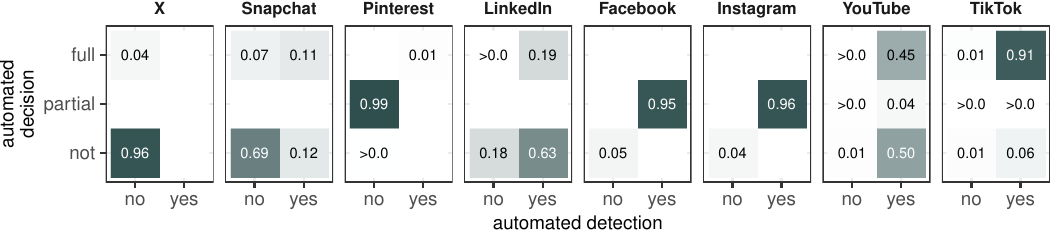} 
\caption{Platform-wise distributions of the use of automated means in statements of reasons, ordered by an ad hoc automation index described by Trujillo et al. \citeyearpar{trujillo2025dsa}.}
\label{fig:sor_automated_means}
\end{figure*}

\paragraph{Regulation}
Article 17(3)(c) states that platforms must detail in their SoRs, where applicable, the use of automated means for content moderation.
Article 42(2)(c) further states that VLOPs must provide indicators of accuracy ``broken down by each official language of the Member States.''

\paragraph{Analysis}
Figure~\ref{fig:sor_automated_means} depicts the joint distribution of automated content detection and automated moderation decisions across platforms.
The figure reveals substantial cross-platform heterogeneity, spanning nearly the entire automation spectrum from predominantly fully automated moderation (TikTok) to almost entirely manual moderation (X). As previously observed for reporting timeliness, Facebook and Instagram exhibit nearly identical patterns, suggesting shared moderation reporting practices or infrastructure. At the same time, several long-standing inconsistencies persist. Most notably, X marked nearly all its SoRs as non-automated despite near-instantaneous moderation, as shown in Appendix Figure~\ref{fig:sor_application_delay}. Additionally, Facebook and Instagram reported extensive use of automated moderation in their \texttt{TRs} but did not classify any SoR as fully automated.

The harmonized \texttt{TRs} also require overall and language-wise classification metrics for automated moderation systems, including \emph{accuracy}, \emph{precision}, and \emph{recall}. Although the EC templates and instructions explicitly specify that these metrics should be reported for all 24 official EU languages using standardized codes, compliance was highly inconsistent. Only Facebook, Instagram, LinkedIn, and X reported language-wise metrics. Instead, Snapchat, TikTok, and YouTube erroneously reported them by country instead of language, and Pinterest did not provide them at all. Moreover, YouTube reported these metrics in two separate files: one for user content moderation and one for ads moderation. Among the four compliant platforms, only Facebook and Instagram covered all 24 official EU languages. LinkedIn covered 22, further indicating that its interface only supports 15 EU languages. X reported metrics for just 7 languages.

The reported classification performance metrics are summarized in Table~\ref{tab:classification_metrics}. As shown, most platforms reported near-perfect performance, with a few notable exceptions concerning \emph{recall}. Specifically, YouTube reported near-zero recall for user content moderation but near-perfect recall for ads moderation, while Facebook and, to a lesser extent, Instagram also reported comparatively low recall values.

\paragraph{Findings}
The reporting of automated moderation remains highly heterogeneous across platforms, with several persistent inconsistencies between SoRs and \texttt{TRs}. Moreover, many platforms failed to comply with the harmonized requirement to report classification metrics by language, thereby limiting the comparability, interoperability, and interpretability of automated moderation data.

\begin{table}[t]
\fontsize{8pt}{10pt}\selectfont
\centering
\begin{tabular*}{0.75\linewidth}{l|rrr}
\toprule
 & \textit{accuracy} & \textit{precision} & \textit{recall} \\ 
\midrule\addlinespace[2.5pt]
Facebook & {\cellcolor[HTML]{FFF5F0}{\textcolor[HTML]{000000}{1.0000}}} & {\cellcolor[HTML]{FFEEE6}{\textcolor[HTML]{000000}{0.9600}}} & {\cellcolor[HTML]{FB7150}{\textcolor[HTML]{FFFFFF}{0.5200}}} \\ 
Instagram & {\cellcolor[HTML]{FFF5F0}{\textcolor[HTML]{000000}{1.0000}}} & {\cellcolor[HTML]{FFF0E9}{\textcolor[HTML]{000000}{0.9700}}} & {\cellcolor[HTML]{FCB599}{\textcolor[HTML]{000000}{0.7300}}} \\ 
LinkedIn & {\cellcolor[HTML]{FFF3EE}{\textcolor[HTML]{000000}{0.9900}}} & {\cellcolor[HTML]{FFF2EB}{\textcolor[HTML]{000000}{0.9800}}} & {\cellcolor[HTML]{FEDFD0}{\textcolor[HTML]{000000}{0.8700}}} \\ 
Pinterest & {\cellcolor[HTML]{FFE8DD}{\textcolor[HTML]{000000}{0.9200}}} & {\cellcolor[HTML]{FFEEE6}{\textcolor[HTML]{000000}{0.9600}}} & {\cellcolor[HTML]{FFE9DF}{\textcolor[HTML]{000000}{0.9300}}} \\ 
Snapchat & {\cellcolor[HTML]{FED6C4}{\textcolor[HTML]{000000}{0.8400}}} & {\cellcolor[HTML]{FFEBE2}{\textcolor[HTML]{000000}{0.9400}}} & {\cellcolor[HTML]{FFEDE4}{\textcolor[HTML]{000000}{0.9500}}} \\ 
TikTok & {\cellcolor[HTML]{FFE7DC}{\textcolor[HTML]{000000}{0.9180}}} & {\cellcolor[HTML]{FFF1EA}{\textcolor[HTML]{000000}{0.9760}}} & {\cellcolor[HTML]{FFEBE1}{\textcolor[HTML]{000000}{0.9380}}} \\ 
X & {\cellcolor[HTML]{FFF3ED}{\textcolor[HTML]{000000}{0.9884}}} & {\cellcolor[HTML]{FFF5F0}{\textcolor[HTML]{000000}{0.9980}}} & {\cellcolor[HTML]{FFF3ED}{\textcolor[HTML]{000000}{0.9865}}} \\ 
YouTube & {\cellcolor[HTML]{FFF3ED}{\textcolor[HTML]{000000}{0.9892}}} & {\cellcolor[HTML]{FEDDCD}{\textcolor[HTML]{000000}{0.8635}}} & {\cellcolor[HTML]{68000D}{\textcolor[HTML]{FFFFFF}{0.0015}}} \\ 
YouTube (ads) & {\cellcolor[HTML]{FFF1EA}{\textcolor[HTML]{000000}{0.9770}}} & {\cellcolor[HTML]{FFEFE7}{\textcolor[HTML]{000000}{0.9627}}} & {\cellcolor[HTML]{FFF5F0}{\textcolor[HTML]{000000}{0.9996}}} \\ 
\bottomrule
\end{tabular*}
\caption{Self-reported overall values for classification metrics used for moderation. YouTube reported separate data for ads, included for completeness due to  differing values.}
\label{tab:classification_metrics}
\end{table}

\subsection{Own-Initiative Actions}
\label{sec:own_initiative}
Own-initiative actions constitute the vast majority of platform moderation interventions and therefore represent a particularly informative setting for assessing DSA transparency and accountability obligations.

\paragraph{Regulation}
Article 17 requires platforms to submit a SoR whenever they impose content moderation restrictions on the grounds that content is illegal or incompatible with their terms and conditions (T{\small\&}C). Recital 54 further clarifies that this obligation also applies to moderation actions taken on the platforms’ own initiative.

\paragraph{Analysis}
Most platforms neither self-reported nor generated SoRs for own-initiative actions explicitly grounded on illegality. Facebook and Instagram were the main exceptions. Several platforms instead appeared to subsume illegal content under violations of their T{\small\&}C, although with heterogeneous reporting practices and explanations. LinkedIn and Pinterest explicitly stated that content violating both the law and platform policies is reported under T{\small\&}C grounds, while Snapchat and TikTok stated that all own-initiative actions are grounded on T{\small\&}C violations. Nevertheless, TikTok still produced over one hundred SoRs for own-initiative illegal content. Snapchat was the only platform to systematically use the \texttt{TDB} field indicating that T{\small\&}C-based actions also involved illegal content, in 28.9\% of cases. By contrast, X stated that it ``does not take measures at it’s [sic] own initiative on the basis of illegality.''

\input{tab-own-initiative}

Table~\ref{tab:own_initiative_automated_prop} compares the number of own-initiative T{\small\&}C moderation actions reported in the \texttt{TDB} and \texttt{TRs}, which should in principle coincide. The log-odds quantify the discrepancy between the two reporting mechanisms, with values near zero indicating close agreement and positive values indicating that platforms reported more actions in the \texttt{TRs} than in the \texttt{TDB}. While Facebook, Instagram, Pinterest, and YouTube ads exhibit strong consistency across the two sources, other platforms show substantial divergences. In particular, LinkedIn, Snapchat, and YouTube reported orders of magnitude more own-initiative actions in their \texttt{TRs} than were observable in the corresponding SoRs. These discrepancies are especially notable because own-initiative T{\small\&}C actions constitute the vast majority of SoRs for most platforms, suggesting persistent inconsistencies between platform self-reporting and \texttt{TDB} records despite harmonization.

\begin{figure}[t]
\centering
\includegraphics[width=1\linewidth]{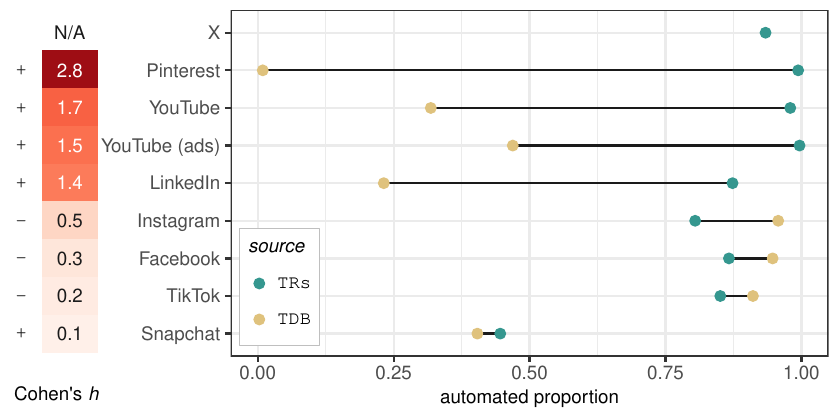} 
\caption{Platform-wise proportions of T{\small\&}C own-initiative actions with solely automated means in the Transparency Database (\texttt{TDB}) and Reports (\texttt{TRs}), sorted by effect size of their difference (Cohen's $h$). Partially and fully automated \texttt{TDB} decisions are grouped as involving automated means.}
\label{fig:own_initiative_automated_prop}
\end{figure}

Finally, we compared the proportion of own-initiative T{\small\&}C actions involving automated means across the \texttt{TDB} and \texttt{TRs}. To enable comparison across the differing reporting granularities of the two mechanisms, we treated partially and fully automated TDB decisions as a single category indicating that automated means were involved. Given the very large sample sizes, all two-sample tests for equality of proportions are statistically significant. We thus measured the effect size of inequality using Cohen's $h$, which ranges from 0 (equal proportions) to the constant $\pi$ (maximum inequality). Figure~\ref{fig:own_initiative_automated_prop} shows the platform-wise magnitude of discrepancies between the two reporting mechanisms. While Facebook, TikTok, and Snapchat exhibit relatively small differences, Pinterest, YouTube, and LinkedIn show very large discrepancies, consistently self-reporting substantially higher levels of automated moderation in their \texttt{TRs} than in the corresponding SoRs. Notably, YouTube ads exhibit near-perfect agreement in action counts between the two sources (log-odds $=-0.015$), but still show a very large discrepancy in the proportion of automated actions ($h=1.5$).

\paragraph{Findings}
Own-initiative T{\small\&}C actions exhibited striking cross-platform and cross-mechanism discrepancies in both action volume and reported use of automated moderation. These inconsistencies indicate persistent interoperability limitations and heterogeneous implementations of DSA reporting obligations despite the harmonized framework.

\subsection{Notice and Action Mechanisms}
\label{sec:notice_and_action_mechanisms}
Unlike own-initiative moderation, notice-based actions originate from external reports, making them a useful test of whether platforms can consistently trace and report the pathway from notice to moderation outcome.

\paragraph{Regulation}
Article 16(1) states that platforms ``shall put mechanisms in place to allow any individual or entity to notify them of the presence on their service of specific items of information that the individual or entity considers to be illegal content.'' Article 22(1) states that priority shall be given to notices of special  entities called \emph{Trusted flaggers}, which have expertise in illegal content. This status is granted by national Digital Services Coordinators. In both the \texttt{TDB} and \texttt{TRs}, notices and actions are distinguished by those initiated via Trusted flaggers and those via Article 16 (i.e., by any other individual or entity).

\paragraph{Analysis}
Similar to the analysis of own-initiative actions, we compared the number of moderation actions initiated via Article 16 notices and Trusted Flaggers across the \texttt{TDB} and the harmonized \texttt{TRs}, which should in principle describe the same activity. Table~\ref{tab:notice_actions_comparison} reports the corresponding counts together with the log-odds discrepancy between the two reporting mechanisms. While some platforms exhibit relatively close agreement, others show substantial divergences, often spanning several orders of magnitude. For Article 16 notices, X self-reported hundreds of thousands of actions despite the complete absence of corresponding SoRs in the \texttt{TDB}. Conversely, Snapchat and LinkedIn reported vastly more notice-based actions in the \texttt{TDB} than in their \texttt{TRs}, whereas Pinterest and YouTube showed the opposite pattern. Discrepancies concerning Trusted Flaggers were similarly heterogeneous. Pinterest self-reported hundreds of Trusted Flagger actions despite having no corresponding SoRs, while YouTube exhibited an extreme mismatch between \texttt{TRs} and \texttt{TDB} counts (84 vs. 1). 

\input{tab-notice-actions}

\paragraph{Findings}
Article 16 and Trusted Flagger actions exhibited substantial inconsistencies across platforms and transparency mechanisms, often involving order-of-magnitude discrepancies between \texttt{TDB} and \texttt{TRs} reporting. Given the central role of notice-based mechanisms in the DSA framework, these stark discrepancies raise serious concerns.

%% file: tab-own-initiative.tex
\begin{table}[t]
    \fontsize{8pt}{10pt}\selectfont
    \begin{tabular*}{\linewidth}{@{\extracolsep{\fill}}lrrrr}
        \toprule
         & \multicolumn{2}{c}{{\texttt{TDB}}} & \multicolumn{2}{c}{{\texttt{TRs}}} \\ 
        \cmidrule(lr){2-3} \cmidrule(lr){4-5}
         & \% of total & \# of actions & \# of actions & log-odds \\ 
        \midrule\addlinespace[2.5pt]
        Facebook & 99.90\% & 183.87M & 171.81M & {\cellcolor[HTML]{F5F4F2}{\textcolor[HTML]{000000}{--0.068}}} \\ 
        Instagram & 99.84\% & 53.09M & 55.92M & {\cellcolor[HTML]{F3F4F4}{\textcolor[HTML]{000000}{+0.052}}} \\ 
        LinkedIn & 82.38\% & 0.15M & 4.01M & {\cellcolor[HTML]{2E8E86}{\textcolor[HTML]{FFFFFF}{+3.299}}} \\ 
        Pinterest & 99.71\% & 168.57M & 184.34M & {\cellcolor[HTML]{F1F4F4}{\textcolor[HTML]{000000}{+0.089}}} \\ 
        Snapchat & 28.20\% & 0.69M & 19.79M & {\cellcolor[HTML]{2B8C84}{\textcolor[HTML]{FFFFFF}{+3.357}}} \\ 
        TikTok & 99.96\% & 169.18M & 258.69M & {\cellcolor[HTML]{E2F1EE}{\textcolor[HTML]{000000}{+0.425}}} \\ 
        X & — & — & 20.87M & {\cellcolor[HTML]{FFFFFF}{\textcolor[HTML]{000000}{—}}} \\ 
        YouTube & 0.32\% & 0.17M & 29.95M & {\cellcolor[HTML]{003D31}{\textcolor[HTML]{FFFFFF}{+5.186}}} \\ 
        YouTube (ads) & 94.41\% & 48.98M & 48.25M & {\cellcolor[HTML]{F5F5F4}{\textcolor[HTML]{000000}{--0.015}}} \\ 
        \bottomrule
    \end{tabular*}
    \caption{Comparison of T{\small\&}C own-initiative actions in the Transparency Database (\texttt{TDB}) and Reports (\texttt{TRs}). For the \texttt{TDB} we show the percentage these actions represent of the platform total. Log-odds quantify the discrepancy between mechanisms: near-zero for close agreement, while positive values indicate more actions in \texttt{TRs} than in the \texttt{TDB}.} 
    \label{tab:own_initiative_automated_prop}
\end{table}

%% file: tab-notice-actions.tex
\begin{table}[t]
    \fontsize{8pt}{10pt}\selectfont
    \setlength{\tabcolsep}{4pt} 
    \begin{tabular*}{\linewidth}{@{\extracolsep{\fill}}lrrrrrr}
        \toprule
         & \multicolumn{3}{c}{{Article 16}} & \multicolumn{3}{c}{{Trusted flaggers}} \\ 
        \cmidrule(lr){2-4} \cmidrule(lr){5-7}
         & \texttt{TDB} &\texttt{TRs} & log-odds & \texttt{TDB} &\texttt{TRs} & log-odds \\ 
        \midrule\addlinespace[2.5pt]
        Facebook & 186,213 & 155,213 & {\cellcolor[HTML]{F5F3EE}{\textcolor[HTML]{000000}{--0.182}}} & 809 & 1,116 & {\cellcolor[HTML]{EAF2F1}{\textcolor[HTML]{000000}{+0.322}}} \\ 
        Instagram & 81,599 & 89,275 & {\cellcolor[HTML]{F2F4F4}{\textcolor[HTML]{000000}{+0.090}}} & 208 & 476 & {\cellcolor[HTML]{D7EEEB}{\textcolor[HTML]{000000}{+0.828}}} \\ 
        LinkedIn & 24,570 & 484 & {\cellcolor[HTML]{BB7D2A}{\textcolor[HTML]{FFFFFF}{--3.927}}} & — & — & {\cellcolor[HTML]{FFFFFF}{\textcolor[HTML]{000000}{—}}} \\ 
        Pinterest & 335 & 1,362 & {\cellcolor[HTML]{C0E7E1}{\textcolor[HTML]{000000}{+1.403}}} & — & 362 & {\cellcolor[HTML]{FFFFFF}{\textcolor[HTML]{000000}{—}}} \\ 
        Snapchat & 1,755,538 & 2,961 & {\cellcolor[HTML]{543005}{\textcolor[HTML]{FFFFFF}{--6.385}}} & 390 & 407 & {\cellcolor[HTML]{F3F5F4}{\textcolor[HTML]{000000}{+0.043}}} \\ 
        TikTok & 59,412 & 95,993 & {\cellcolor[HTML]{E4F1EF}{\textcolor[HTML]{000000}{+0.480}}} & 163 & 120 & {\cellcolor[HTML]{F6F2E9}{\textcolor[HTML]{000000}{--0.306}}} \\ 
        X & — & 388,095 & {\cellcolor[HTML]{FFFFFF}{\textcolor[HTML]{000000}{—}}} & — & 922 & {\cellcolor[HTML]{FFFFFF}{\textcolor[HTML]{000000}{—}}} \\ 
        YouTube & 210,933 & 418,761 & {\cellcolor[HTML]{DDEFEC}{\textcolor[HTML]{000000}{+0.686}}} & 1 & 84 & {\cellcolor[HTML]{208078}{\textcolor[HTML]{FFFFFF}{+4.431}}} \\ 
        \bottomrule
    \end{tabular*}
    \caption{Count comparison of actions in the Transparency Database (\texttt{TDB}) and Reports (\texttt{TRs}) initiated via either \emph{Article 16} or \emph{Trusted flaggers} notices. Log-odds quantify the discrepancy between the two reporting mechanisms.}
    \label{tab:notice_actions_comparison}
\end{table}

%% file: 5-evaluation.tex
\section{Comparative Assessment}
\label{sec:assessment}
To synthesize the empirical results into a comparable and actionable assessment, we conducted a structured comparison of reporting quality across platforms and transparency mechanisms based on the previous quantitative analyses as well as the ones reported in the Appendix.

\subsection{Assessment Framework}
We assess reporting quality along four dimensions, according to their applicability to each mechanism: timeliness (\texttt{TDB}); consistency (\texttt{TDB}~{\small\&}~\texttt{TRs}); completeness (\texttt{TDB}~{\small\&}~\texttt{TRs}); and formatting (\texttt{TRs}). For each applicable dimension, the assessment summarizes the preceding quantitative evidence through three ordinal adequacy levels: \emph{adequate}, when there is no identified issue, or there are only minor issues that do not prevent the reported information from supporting evaluation of the dimension; \emph{inadequate}, when identifiable omissions, inconsistencies, or other reporting issues prevent evaluation of some essential aspects of the dimension; and \emph{semi-adequate}, otherwise, when certain identifiable issues exist, but do not affect the reported information in ways that preclude evaluation of the essential aspects of the dimension. We deliberately keep the four dimensions separate, as they capture complementary properties of reporting quality and jointly provide a multidimensional assessment of each platform and transparency mechanism. Moreover, the relative importance of any one dimension may depend on specific analytical questions or the normative context.

\input{tab-ordinal-scores}

\emph{Timeliness} evaluates whether SoRs are submitted to the \texttt{TDB} without substantial delays or prolonged interruptions, based on the patterns identified in the submission time series. This criterion only applies to the \texttt{TDB}, since SoRs are timestamped and intended to be submitted ``without undue delay,'' whereas \texttt{TRs} are periodic aggregate disclosures for which fine-grained reporting latency is largely immaterial. \emph{Consistency} evaluates the extent to which reported data are internally coherent and interoperable between transparency mechanisms when subject to basic cross-checks, including agreement in action counts, moderation categories, and use of automated means. \emph{Completeness} evaluates whether the required information is sufficiently present and detailed to support meaningful analysis and comparison, including the presence of mandatory attributes, language-wise reporting, and sufficiently granular disclosures. Finally, \emph{formatting} evaluates the extent to which harmonized \texttt{TRs} comply with the EC templates, specifications, and instructions. This criterion only applies to \texttt{TRs}, since the \texttt{TDB} already enforces a fixed submission schema at the database level. Assessments were assigned based on the empirical patterns, inconsistencies, omissions, and reporting limitations identified throughout the preceding quantitative analyses.

Importantly, limitations across these dimensions have different implications for the use of transparency data. Formatting inconsistencies, while costly to process, primarily affect machine readability and processing, and can sometimes be corrected during analysis, as we also did throughout the previous quantitative analyses to enable several systematic comparisons that would be otherwise impossible. Timeliness issues affect the availability of information for timely oversight, while consistency issues affect the reliability and interoperability of corresponding disclosures. Completeness issues constrain the scope of possible analyses because missing information cannot generally be reconstructed from the reported data. These distinct implications further motivate assessing the dimensions separately rather than treating reporting quality as a single property.

Table~\ref{tab:platform_evaluation} presents the resulting structured expert synthesis, providing a compact overview of reporting adequacy across platforms, mechanisms, and dimensions. To illustrate how the assessment criteria translate into individual assignments, consider \emph{timeliness}. LinkedIn represents a clear case of \emph{adequate} reporting: it submitted SoRs every day throughout the reporting period and exhibited an almost-zero median communication delay. Conversely, X represents a clear case of \emph{inadequate} reporting, as it featured 52 days without SoR submissions (28.2\% of the reporting period), with delayed submissions extending over approximately three months and a median communication delay of 30.9 days. Pinterest illustrates the intermediate \emph{semi-adequate} category. It experienced a 16-day interruption in SoR submissions, but subsequently recovered while maintaining an almost-zero median communication delay.
For transparency and traceability of the full assessment, Appendix Table~\ref{tab:platform_evaluation_commented} reports the empirical observations underlying every individual assignment in Table~\ref{tab:platform_evaluation}. The following subsection examines the most relevant platform-specific cases underlying these assessments, after which we synthesize cross-platform patterns and broader insights emerging from the comparative analysis.

\subsection{Platform-Level Assessment}
\paragraph{Facebook and Instagram}
Both Meta platforms share the same internal guidelines, procedures, infrastructure, and human resources for content moderation. Accordingly, their reporting quality is also nearly identical. We nevertheless observed a few differences. For instance, Instagram exhibited Cyprus as a suspicious normalized AMAR outlier, whereas this was less marked for Facebook. Instagram also showed a relatively larger discrepancy concerning automated means in own-initiative actions. Incidentally, most Meta SoRs (more than 75\%) are associated with the content type \emph{other}, although these cases are exclusively labeled as \emph{accounts}. Additional issues include a hiccup in SoR submissions, the absence of fully automated SoRs despite Meta reporting extensive automated moderation in its \texttt{TRs}, and the use of custom \emph{other} infringement categories instead of the dedicated free-text reporting field. Nevertheless, both platforms exhibited relatively limited discrepancies between the \texttt{TDB} and \texttt{TRs} concerning notice and own-initiative actions. In addition, the qualitative sections and contextual information provided in the \texttt{TRs} are generally informative and appropriate.

\paragraph{LinkedIn}
LinkedIn presented relatively few but important irregularities. Most notably, there are substantial discrepancies between the number of actions reported in the \texttt{TDB} and \texttt{TRs}. Notice-based actions are considerably more numerous in the \texttt{TDB}, whereas own-initiative T{\small\&}C actions are far more prevalent in the \texttt{TRs} and also exhibit markedly different proportions of automated processing. Another issue is the systematic use of the content type \emph{other} in SoRs, together with inconsistent descriptive labels for these entries. The platform also reported classification metrics by country rather than by language, contrary to the EC reporting specifications.

\paragraph{Pinterest}
This pinboard platform reported the content type of all its SoRs as \emph{other}, with these being overwhelmingly (99.7\%) labeled as \emph{Pin}, despite that they can refer to an image, video, or rich media. It also had a significant gap in the submission of SoRs, but it was able to recover and keep a low communication delay. It was the only platform that did not provide classification metrics by neither country nor language, most likely due to the visual nature of its content. Concerning discrepancies between mechanisms, it had a relatively low difference for own-initiative actions but a very large discrepancy of solely automated proportion (the highest). Perplexingly, Pinterest had circa 2 SoRs per AMAR at the EU level, which is double the value for TikTok (the second largest). For reference, for its initial \texttt{TDB} data Pinterest had circa 0.5 SoRs per AMAR~\citepalias{trujillo2025dsa}. This striking increment seems odd because, whereas TikTok mostly processes SoRs with fully automated means (91\%), Pinterest overwhelmingly uses non-automated detection with partially automated decision (99\%). It should be noted, however, that the platform self-reports very high levels of solely automated means. Curiously, the time difference between content creation and decision application for SoRs had the highest average (median of 372 days). The platform also had a large discrepancy for actions via notices, with significantly fewer or missing SoRs for Article 16 and trusted flaggers, respectively.

\paragraph{Snapchat}
Another platform that presented few but important irregularities. In particular, there were remarkable discrepancies between the \texttt{TDB} and \texttt{TRs} in the volume of actions, both by own initiative (much higher \texttt{TRs} volume) and  notices (strikingly higher \texttt{TDB} volume). On the other hand,
the \texttt{TRs} quality in formatting is high and the given contextual information is pertinent.
For example, for automated means it indicated that it does not generally track the language of content or that this is not possible due to content type, hence it reported data at the country level as a proxy for language-information.
Another such contextual information is that all trusted flagger notices are reviewed without automated means.
Additionally, and as mentioned before, Snapchat was the only platform to correctly mark SoRs that are violations against both the law and T{\small\&}C.

\paragraph{TikTok}
This short-video platform had very noticeable issues with SoR submissions and communication delays throughout the reporting period. There were rather considerable discrepancies in the volume of actions between SoRs and reports---initiated both by notices or own initiative---albeit the solely automated proportion for the latter is relatively consistent. One peculiar setback we encountered was the use of ISO country codes for 7 of the 24 EU languages in the file for moderation human resources---e.g., \texttt{ie} (Ireland) instead of \texttt{ga} (Gaeilge). We deduced these indeed referred to languages based on the values of the other records and related qualitative reporting. Still, TikTok was one of the three platforms that reported classification metrics by country and not language. On a more positive note, TikTok was the platform with more complete SoRs. For instance, it had the widest use and distribution in content type, although \emph{synthetic media} was never used and the descriptive labeling for those of content type \emph{other} (8\%) was rather inconsistent.

\paragraph{X}
This micro-blogging platform presented contrasting evaluations for both reporting mechanisms. On one hand, it had several and important issues concerning its data on the \texttt{TDB}. As already identified by Trujillo et al.~\citeyearpar{trujillo2025dsa}, since the start of the database X submits SoRs exclusively on illegal content (initiated via unspecified notices), but with several critical inconsistencies. For example, almost all SoRs (218K) are \emph{synthetic media}, with just a few dozens being \emph{text} and a handful \emph{audio}. Moreover, practically all SoRs are processed without any automation the same day in which the content was created, which is implausible. Further still, there was a remarkable lack of submissions in the first part of the period of study and high communication delays afterward. On the other hand, the \texttt{TRs} by X offered much more detailed, consistent, and complete information. This strong mismatch between approaches and quality of reporting seem to indicate that independent teams are in charge of each transparency mechanism. Despite their merits, there were a few light issues with the \texttt{TRs}, such as providing a slightly higher number of moderators proficient in English than the reported total, as shown in Appendix Figure~\ref{fig:language_moderators}.

\paragraph{YouTube}
The most popular platform had several issues with report formatting. First, it provided separate data files for ads in three sections of \texttt{TRs}. Then, it provided ranges of values instead of their averages in automated means. Finally and most importantly, for self-reported AMAR, it only provided the URL of its PDF report instead of directly reporting the data as requested. This last issue goes against the very purpose of the harmonization of \texttt{TRs}, which prescribes a machine-readable format to ease processing and analysis. 
Further still, AMAR data presented remarkable irregularities in which some countries had much more active recipients than a reasonable upper limit, which is more generous than the one indicated by Google itself. More substantively, YouTube reported an extreme discrepancy in classification recall between user-content and ads moderation (0.0015 vs.~0.9996), raising questions about the comparability and interpretation of the reported automated-moderation metrics. For SoRs, we also found striking discrepancies for own-initiative actions on user content and notice actions of both kinds, in addition to large differences between solely automated own-initiative actions for both ads and user content.

\subsection{Cross-Platform Insights}
The main insight emerging from the comparative assessment is that transparency reporting quality is inherently multidimensional. No platform is uniformly adequate across all applicable dimensions, and performance along one dimension does not necessarily extend to others. For example, TikTok combines inadequate \texttt{TDB} timeliness with adequate completeness, while YouTube combines adequate \texttt{TDB} timeliness and completeness with inadequate \texttt{TR} formatting. Differences can also emerge between transparency mechanisms within the same platform, most prominently for X. More generally, Table~\ref{tab:platform_evaluation} reveals that recurring cross-platform limitations coexist with platform- and mechanism-specific weaknesses. Consequently, neither compliance nor transparency reporting quality can be meaningfully reduced to a single overall assessment: timeliness, consistency, completeness, and formatting capture distinct properties of reporting and should be evaluated separately.

%% file: tab-ordinal-scores.tex
\begin{table}[t]
\centering
\fontsize{8pt}{10pt}\selectfont
\begin{tabular*}{0.7\linewidth}{lcccccc}
\toprule
 & \multicolumn{3}{c}{{\texttt{TDB}}} & \multicolumn{3}{c}{{\texttt{TRs}}} \\ 
\cmidrule(lr){2-4} \cmidrule(lr){5-7}
 & \emph{tm} & \emph{cn} & \emph{cp} & \emph{fm} & \emph{cn} & \emph{cp} \\
\midrule\addlinespace[2.5pt]
Facebook & \tikz[baseline=-0.6ex] \fill[semiadequate] circle (0.6ex); & \tikz[baseline=-0.6ex] \fill[semiadequate] circle (0.6ex); & \tikz[baseline=-0.6ex] \fill[semiadequate] circle (0.6ex); & \tikz[baseline=-0.6ex] \fill[semiadequate] circle (0.6ex); & \tikz[baseline=-0.6ex] \fill[semiadequate] circle (0.6ex); & \tikz[baseline=-0.6ex] \fill[adequate] circle (0.5ex); \\ 
Instagram & \tikz[baseline=-0.6ex] \fill[semiadequate] circle (0.6ex); & \tikz[baseline=-0.6ex] \fill[semiadequate] circle (0.6ex); & \tikz[baseline=-0.6ex] \fill[semiadequate] circle (0.6ex); & \tikz[baseline=-0.6ex] \fill[semiadequate] circle (0.6ex); & \tikz[baseline=-0.6ex] \fill[semiadequate] circle (0.6ex); & \tikz[baseline=-0.6ex] \fill[adequate] circle (0.5ex); \\ 
LinkedIn & \tikz[baseline=-0.6ex] \fill[adequate] circle (0.5ex); & \tikz[baseline=-0.6ex] \fill[semiadequate] circle (0.6ex); & \tikz[baseline=-0.6ex] \fill[semiadequate] circle (0.6ex); & \tikz[baseline=-0.6ex] \fill[adequate] circle (0.5ex); & \tikz[baseline=-0.6ex] \fill[semiadequate] circle (0.6ex); & \tikz[baseline=-0.6ex] \fill[adequate] circle (0.5ex); \\ 
Pinterest & \tikz[baseline=-0.6ex] \fill[semiadequate] circle (0.6ex); & \tikz[baseline=-0.6ex] \fill[semiadequate] circle (0.6ex); & \tikz[baseline=-0.6ex] \fill[semiadequate] circle (0.6ex); & \tikz[baseline=-0.6ex] \fill[adequate] circle (0.5ex); & \tikz[baseline=-0.6ex] \fill[adequate] circle (0.5ex); & \tikz[baseline=-0.6ex] \fill[semiadequate] circle (0.6ex); \\ 
Snapchat & \tikz[baseline=-0.6ex] \fill[adequate] circle (0.5ex); & \tikz[baseline=-0.6ex] \fill[semiadequate] circle (0.6ex); & \tikz[baseline=-0.6ex] \fill[semiadequate] circle (0.6ex); & \tikz[baseline=-0.6ex] \fill[semiadequate] circle (0.6ex); & \tikz[baseline=-0.6ex] \fill[semiadequate] circle (0.6ex); & \tikz[baseline=-0.6ex] \fill[semiadequate] circle (0.6ex); \\ 
TikTok & \tikz[baseline=-0.6ex] \fill[inadequate] circle (0.75ex); & \tikz[baseline=-0.6ex] \fill[semiadequate] circle (0.6ex); & \tikz[baseline=-0.6ex] \fill[adequate] circle (0.5ex); & \tikz[baseline=-0.6ex] \fill[semiadequate] circle (0.6ex); & \tikz[baseline=-0.6ex] \fill[adequate] circle (0.5ex); & \tikz[baseline=-0.6ex] \fill[adequate] circle (0.5ex); \\ 
X & \tikz[baseline=-0.6ex] \fill[inadequate] circle (0.75ex); & \tikz[baseline=-0.6ex] \fill[inadequate] circle (0.75ex); & \tikz[baseline=-0.6ex] \fill[inadequate] circle (0.75ex); & \tikz[baseline=-0.6ex] \fill[adequate] circle (0.5ex); & \tikz[baseline=-0.6ex] \fill[semiadequate] circle (0.6ex); & \tikz[baseline=-0.6ex] \fill[semiadequate] circle (0.6ex); \\ 
YouTube & \tikz[baseline=-0.6ex] \fill[adequate] circle (0.5ex); & \tikz[baseline=-0.6ex] \fill[semiadequate] circle (0.6ex); & \tikz[baseline=-0.6ex] \fill[adequate] circle (0.5ex); & \tikz[baseline=-0.6ex] \fill[inadequate] circle (0.75ex); & \tikz[baseline=-0.6ex] \fill[semiadequate] circle (0.6ex); & \tikz[baseline=-0.6ex] \fill[semiadequate] circle (0.6ex); \\ 
\bottomrule
\end{tabular*}
\caption{Structured comparative assessment of platform reporting across the Transparency Database (\texttt{TDB}) and Transparency Reports (\texttt{TRs}). The assessment considers timeliness (\emph{tm}), consistency (\emph{cn}), completeness (\emph{cp}), and formatting (\emph{fm}), according to their applicability to each mechanism. Colored dots represent  assessment: \emph{inadequate}~(\tikz[baseline=-0.6ex] \fill[inadequate] circle (0.7ex);), \emph{semi-adequate}~(\tikz[baseline=-0.6ex] \fill[semiadequate] circle (0.5ex);), and \emph{adequate}~(\tikz[baseline=-0.6ex] \fill[adequate] circle (0.4ex);). The empirical evidence underlying each assignment is reported in Appendix Table~\ref{tab:platform_evaluation_commented}.
}
\label{tab:platform_evaluation}
\end{table}

%% file: 6-discussion.tex
\section{Discussion and Conclusions}
\label{sec:discussion}

\subsection{Lack of Harmonization and Reliability}

\paragraph{RQ1: Compliance with Instructions}
All eight platforms exhibited non-trivial compliance issues with respect to the EC specifications and reporting instructions, across both the \texttt{TDB} and harmonized \texttt{TRs}. Importantly, the observed non-compliance varied not only in magnitude across platforms, but also in nature. At the syntactic level, most reporting specifications appeared sufficiently clear, yet platforms frequently failed to follow them consistently. Examples include the incorrect use of language codes, incomplete contextual information, and failure to properly flag actions involving both legal and T{\small\&}C violations. At the semantic level, however, the harmonization effort exposed more fundamental divergences consistent with differences in how platforms interpreted and operationalized the reporting obligations. This was particularly evident for concepts such as automated moderation, granular language-wise reporting, and the treatment of distinct moderation domains such as advertising. On the one hand, these findings signal that the EC might need to revise parts of the current reporting mechanisms and provide further reporting guidance. On the other hand, they suggest that the current harmonization framework improved structural standardization, but did not fully achieve semantic consistency or operational interoperability across platforms and reporting mechanisms.

\paragraph{RQ2: Consistency between Mechanisms}
All eight platforms exhibited substantial inconsistencies between the \texttt{TDB} and harmonized \texttt{TRs}. Overall, reporting data in the \texttt{TDB} were generally less complete and internally consistent than the corresponding information in the \texttt{TRs}, despite both mechanisms being intended to describe the same moderation activity. These discrepancies are consistent with partially independent reporting workflows, limited internal interoperability, or heterogeneous interpretations of the DSA obligations, although the underlying causes cannot be established from the reported data alone. A prominent example was X, for which we identified substantial reporting issues in the \texttt{TDB} despite considerably fewer issues in the harmonized \texttt{TRs}. More broadly, our findings indicate that post-harmonization transparency reporting still lacks sufficient operational consistency across mechanisms, thereby limiting the reliability, comparability, and auditability of DSA transparency data.

\paragraph{} 
Taken together, these findings suggest that scholars and practitioners should interpret DSA transparency data with caution, particularly when drawing comparative or large-scale conclusions without additional validation and cross-checking procedures.

\subsection{Suggestions to Improve Reporting Mechanisms}
Our findings suggest that several targeted modifications to the DSA transparency infrastructure could substantially improve the interoperability, consistency, and auditability of transparency reporting. Notably, the harmonization changes introduced for the 2025H2 reporting period primarily aligned reporting structures and categories between the \texttt{TDB} and \texttt{TRs}, while leaving several previously identified operational limitations unresolved. In this regard, our analyses further support prior proposals aimed at improving the \texttt{TDB}, including making more attributes mandatory (especially those concerning moderation grounds), explicitly distinguishing moderation targets from the entities causing the violation (e.g., content versus accounts), separately identifying synthetic or AI-generated targets, and clarifying the semantics of automated moderation attributes~(\citealt{kaushal2024automated}; \citetalias{trujillo2025dsa}). Our results also highlight the need for additional improvements specifically concerning harmonized \texttt{TRs}:
\begin{itemize}
    \item \textit{Prescriptive semantic guidance.} Several inconsistencies cannot be explained by formatting differences alone and are consistent with heterogeneous interpretations of the same reporting obligations. This was particularly evident for concepts such as solely, partially, and non-automated moderation. More explicit operational definitions and reporting guidelines would therefore improve semantic consistency and cross-platform comparability.
    \item \textit{Constrained extensibility.} Multiple platforms were unable to provide some requested information in the prescribed format and instead reported alternative representations, such as country-wise rather than language-wise data, or separate reporting structures for distinct moderation domains (e.g., YouTube ads). Rather than forcing these cases into rigid templates, the reporting framework should support controlled extensions under clearly specified conditions and metadata requirements.
    \item \textit{Automated validation.} Several platforms did not fully comply with the EC templates and reporting instructions, substantially complicating data processing and interoperability. Harmonized \texttt{TRs} should therefore undergo mandatory automated validation against formal schemas and reporting constraints before publication. Such validation could additionally include integrity-preserving mechanisms, such as metadata registration and cryptographic hashing of submitted files. Beyond improving interoperability and reproducibility, this would likely eliminate many of the formatting issues identified herein.
\end{itemize}

\subsection{Policy Implications for Moderation Transparency}
Our findings have implications not only for DSA transparency reporting mechanisms, but also for the broader governance of online content moderation transparency. At the platform level, the observed discrepancies suggest the need for stronger internal coordination between reporting procedures, moderation infrastructures, and transparency workflows. While our data do not reveal the internal processes producing these discrepancies, the observed inconsistencies are compatible with heterogeneous operational interpretations of the same regulatory obligations across different reporting mechanisms within the same platform. At the regulatory level, our results highlight that formal harmonization alone is insufficient to guarantee interoperability, consistency, or auditability of transparency data. Effective transparency regulation additionally requires continuous validation procedures, operational guidance, and meaningful enforcement of reporting obligations. Our findings also distinguish between issues that appear tractable through enforcement of existing requirements and those demanding further regulatory intervention. Clear deviations from prescribed templates and reporting instructions, such as incorrect language-wise reporting or missing required information, can be addressed through stricter validation and enforcement of existing obligations. Conversely, discrepancies arising from ambiguous definitions, heterogeneous interpretations, or reporting cases not adequately represented by the prescribed templates, require clearer operational guidance or revisions to the reporting framework. This is particularly relevant for VLOPs, whose reporting practices directly affect the transparency and accountability of some of the largest online moderation systems operating in the EU. More broadly, our findings suggest that transparency infrastructures can become formally standardized while remaining operationally inconsistent in practice. This risks undermining one of the central objectives of the DSA: enabling a transparent, predictable, and trustworthy online environment through harmonized accountability mechanisms. Ensuring the reliability and interoperability of transparency data is therefore not merely a technical reporting issue, but a necessary condition for effective public oversight, independent auditing, and regulatory enforcement.

\subsection{Limitations and Future Work}
Our analyses primarily focus on operational consistency, interoperability, and compliance with DSA reporting obligations rather than on modeling the substantive dynamics of content moderation itself. Accordingly, we relied mainly on transparent and interpretable auditing procedures, which proved sufficient to expose substantial reporting inconsistencies across platforms and mechanisms. Our cross-mechanism comparisons assume that corresponding \texttt{TDB} and \texttt{TR} fields describe sufficiently overlapping moderation activities to permit meaningful comparison. However, platforms may operationalize nominally corresponding reporting categories differently. Consequently, observed discrepancies establish inconsistencies between the two mechanisms but do not, by themselves, identify their causes. They may reflect reporting errors, differences in regulatory interpretation, aggregation or counting procedures, temporal boundaries, technical reporting pipelines, or other implementation choices that cannot be distinguished from the reported data alone. Moreover, as our analyses rely on platform-reported data, they cannot identify moderation actions consistently omitted from both reporting mechanisms or establish which mechanism more accurately reflects underlying moderation activity. Finally, our observational design evaluates the implementation and operational outcomes of harmonization but does not warrant causal identification of either its effects or the mechanisms underlying the observed discrepancies. Finally, our study focuses on the eight originally designated VLOPs and therefore may not fully capture reporting practices across the broader platform ecosystem. However, these platforms represent some of the most influential online services subject to the DSA and are consequently among the most consequential cases for evaluating the effectiveness of harmonized transparency reporting.

%% file: a-appendix.tex

\onecolumn
\section{Appendix}
In this section we briefly depict ---via tables and figures--- additional aspects of the analyses we conducted on the data from both the \texttt{TDB} and \texttt{TRs}.

\vspace{1.5cm} 

\begin{table*}[h]
\centering
\fontsize{8pt}{10pt}\selectfont
\begin{tabular*}{0.35\linewidth}{@{\extracolsep{\fill}}lrrr}
\toprule
 & \#SoRs & AMAR & $\frac{\text{\#SoRs}}{\text{AMAR}}$ \\ 
\midrule\addlinespace[2.5pt]
Pinterest & 169.07M & 84.63M & 1.9978 \\ 
TikTok & 169.25M & 178.30M & 0.9492 \\ 
Facebook & 184.06M & 262.48M & 0.7012 \\ 
Instagram & 53.17M & 288.74M & 0.1842 \\ 
YouTube & 51.88M & 479.40M & 0.1082 \\ 
Snapchat & 2.45M & 97.15M & 0.0252 \\ 
X & 0.22M & 64.77M & 0.0034 \\ 
LinkedIn & 0.18M & 55.20M & 0.0033 \\ 
\bottomrule
\end{tabular*}
\caption{Platform-wise totals for statements of reasons (SoRs) and average monthly active recipients (AMAR).}
\label{tab:sor_per_amar}
\end{table*}

\vspace{1.5cm}

\begin{figure*}[h]
\centering
\includegraphics[width=0.5\linewidth]{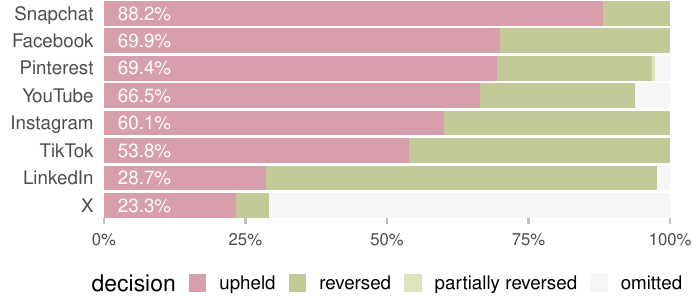} 
\caption{Platform-wise distributions of self-reported decision appeals on internal-complaint mechanisms. Decisions \emph{omitted} refers to appeals that did not lead to a subsequent decision (e.g., withdrawn appeal).  Only Pinterest reported decisions \emph{partially reversed}  (0.5\%).}
\label{fig:internal_complaints_mechanism_decisions}
\end{figure*}

\vspace{1.5cm}

\begin{figure*}[h]
\centering
\includegraphics[width=.9\textwidth]{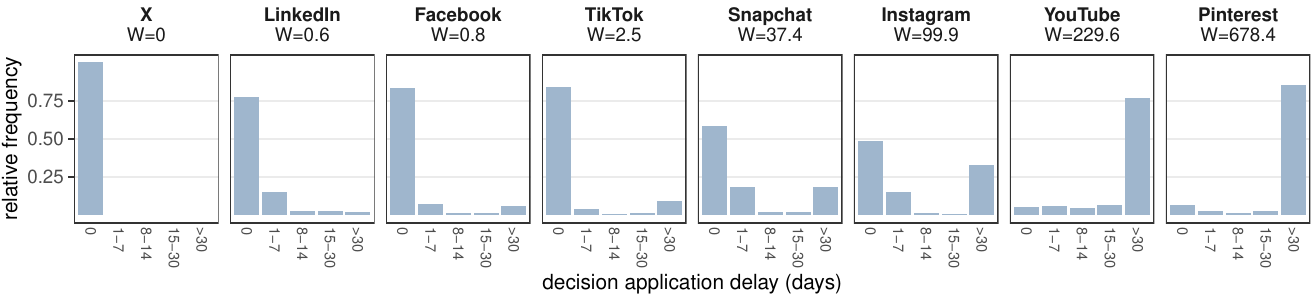} 
\caption{Platform-wise distributions of decision application delays in statements of reasons (compared to the content creation date), sorted by their 80\% winsorized means (W).}
\label{fig:sor_application_delay}
\end{figure*}

\begin{figure*}[t]
\centering
\subcaptionbox{Platform-wise category distribution}{\includegraphics[width=0.485\textwidth]{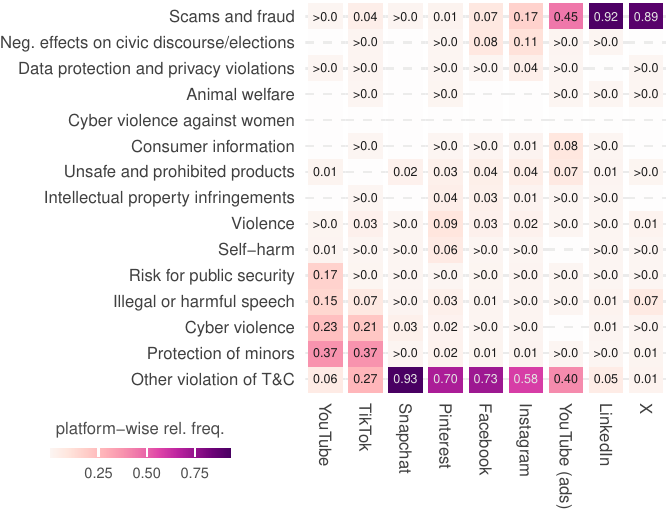}}
\hspace{1.05em}
\subcaptionbox{Processed solely via automated means}{\includegraphics[width=0.485\textwidth]{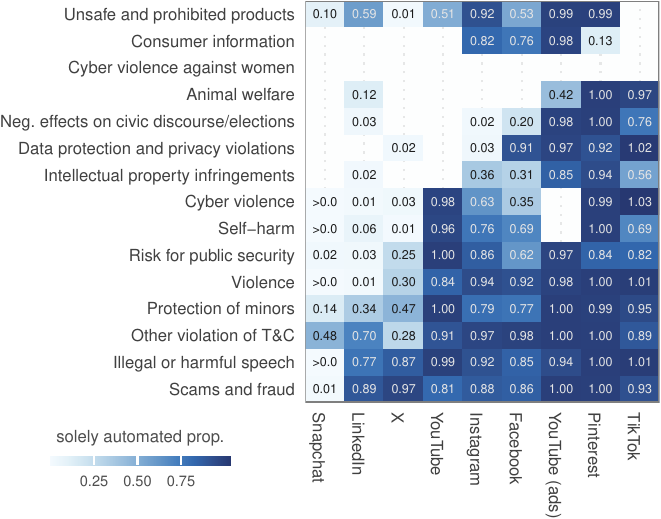}}
\caption{Self-reported own-initiative actions for violations of terms and conditions (T{\small\&}C) by category.  Platforms and categories are ordered by similarity.  YouTube reported separate data for ads, included for completeness.}
\label{fig:own_initiative_prop}
\end{figure*}

\begin{figure*}[t]
\centering
\includegraphics[width=1\linewidth]{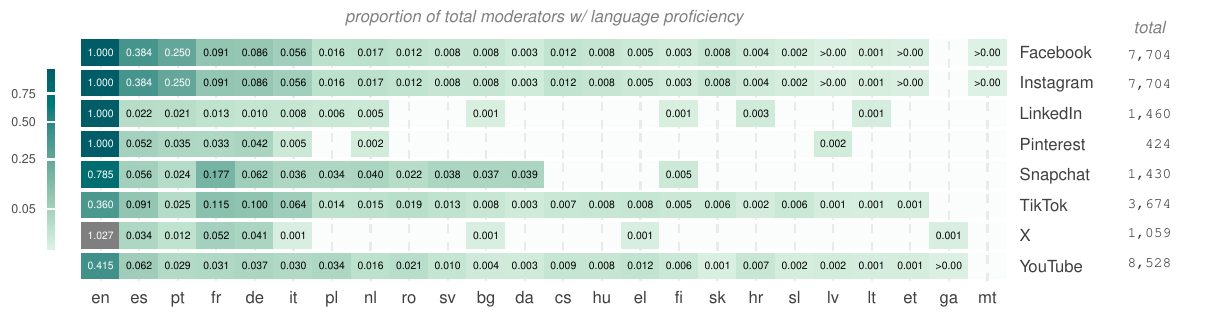}
\caption{Platform-wise self-reported human resources working on moderation by language and total. A moderator might be counted multiple times if proficient in multiple languages. X erroneously set more moderators proficient in English than their reported total. Facebook and Instagram share moderators provided by Meta, which reported all as proficient in English (with 4\% working exclusively in it), and also uses moderators in certain languages (e.g., Spanish, Portuguese) for content outside the EU. YouTube reported 6,024 language-agnostic moderators but did not provide contextual information for these.}
\label{fig:language_moderators}
\end{figure*}

\begin{figure*}[t]
\centering
\includegraphics[width=1\linewidth]{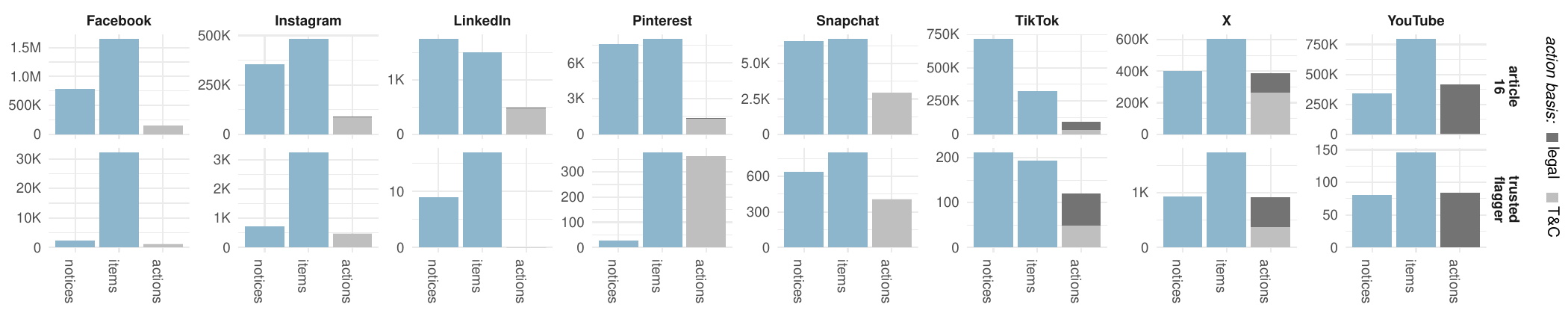}
\caption{Platform-wise self-reported number of received notices ---via either Trusted Flaggers or Article 16 notice and action mechanisms--- their respective number of specific items of information (one notice might refer to multiple items or vice versa), and subsequent moderation actions ---on the basis of legality or violation of terms and conditions (T{\small\&}C).}
\label{fig:notices_flow}
\end{figure*}

\setlength{\tabcolsep}{0.25em}
\begin{table}[t]
\fontsize{7pt}{9pt}\selectfont
\begin{tabular*}{\linewidth}{@{\extracolsep{\fill}}l|lll}
\toprule
 & \textbf{platform} &  & \textbf{empirical evidence} \\ 
\midrule\addlinespace[2.5pt]
\multirow{8}{*}{\shortstack[c]{\texttt{TRs}\\\emph{cp}}} & Facebook & \tikz[baseline=-0.6ex] \fill[adequate] circle (0.5ex); & No substantial missing data. \\ 
 & Instagram & \tikz[baseline=-0.6ex] \fill[adequate] circle (0.5ex); & No substantial missing data. \\ 
 & LinkedIn & \tikz[baseline=-0.6ex] \fill[adequate] circle (0.5ex); & Remarkably fewer Art. 16 actions compared to TDB. \\ 
 & Pinterest & \tikz[baseline=-0.6ex] \fill[semiadequate] circle (0.6ex); & Lack accuracy metrics by language or country. \\ 
 & Snapchat & \tikz[baseline=-0.6ex] \fill[semiadequate] circle (0.6ex); & Accuracy metrics by country, not language. \\ 
 & TikTok & \tikz[baseline=-0.6ex] \fill[adequate] circle (0.5ex); & No substantial missing data. \\ 
 & X & \tikz[baseline=-0.6ex] \fill[semiadequate] circle (0.6ex); & Accuracy metrics by country and not language. \\ 
 & YouTube & \tikz[baseline=-0.6ex] \fill[semiadequate] circle (0.6ex); & Link to a PDF report given for AMAR. Missing `total' for language moderators and classification metrics of ads. Accuracy metrics by country, not language. \\ 
\midrule\addlinespace[2.5pt]
\multirow{8}{*}{\shortstack[c]{\texttt{TRs}\\\emph{cn}}} & Facebook & \tikz[baseline=-0.6ex] \fill[semiadequate] circle (0.6ex); & Moderately fewer Art. 16 actions compared to TDB. \\ 
 & Instagram & \tikz[baseline=-0.6ex] \fill[semiadequate] circle (0.6ex); & Moderately fewer Art. 16 actions compared to TDB. \\ 
 & LinkedIn & \tikz[baseline=-0.6ex] \fill[semiadequate] circle (0.6ex); & Remarkably fewer Art. 16 actions compared to TDB. \\ 
 & Pinterest & \tikz[baseline=-0.6ex] \fill[adequate] circle (0.5ex); & No substantial discrepancies. \\ 
 & Snapchat & \tikz[baseline=-0.6ex] \fill[semiadequate] circle (0.6ex); & Remarkably fewer Art. 16 actions compared to TDB. \\ 
 & TikTok & \tikz[baseline=-0.6ex] \fill[adequate] circle (0.5ex); & No substantial discrepancies. \\ 
 & X & \tikz[baseline=-0.6ex] \fill[semiadequate] circle (0.6ex); & Number  moderators for Engish higher than the total for all languages. \\ 
 & YouTube & \tikz[baseline=-0.6ex] \fill[semiadequate] circle (0.6ex); & Highly contrasting classification recall values for ads and user content. \\ 
\midrule\addlinespace[2.5pt]
\multirow{8}{*}{\shortstack[c]{\texttt{TRs}\\\emph{fm}}} & Facebook & \tikz[baseline=-0.6ex] \fill[semiadequate] circle (0.6ex); & Custom subcategories for `Other' instead of using description as instructed. \\ 
 & Instagram & \tikz[baseline=-0.6ex] \fill[semiadequate] circle (0.6ex); & Custom subcateogries for `Other' instead of using description as instructed. \\ 
 & LinkedIn & \tikz[baseline=-0.6ex] \fill[adequate] circle (0.5ex); & Few uses of `N/A' instead of leaving blank values as instructed. \\ 
 & Pinterest & \tikz[baseline=-0.6ex] \fill[adequate] circle (0.5ex); & Wrong country code:  `GR' instead of `EL'. \\ 
 & Snapchat & \tikz[baseline=-0.6ex] \fill[semiadequate] circle (0.6ex); & Percentages instead of values from 0 to 1 for automated means. Wrong country code:  `GR' instead of `EL'. \\ 
 & TikTok & \tikz[baseline=-0.6ex] \fill[semiadequate] circle (0.6ex); & Wrong  code for 7 languages (e.g, `ie' for `ga', `gr' for `el'). Wrong country code:  `GR' instead of `EL'. \\ 
 & X & \tikz[baseline=-0.6ex] \fill[adequate] circle (0.5ex); & Wrong country code:  `GR' instead of `EL'. \\ 
 & YouTube & \tikz[baseline=-0.6ex] \fill[inadequate] circle (0.75ex); & Three reporting  sections have additional files for ads. Accuracy metrics are intervals (really close in value) and not single numbers. Missing AMAR. \\ 
\midrule\addlinespace[2.5pt]
\multirow{8}{*}{\shortstack[c]{\texttt{TDB}\\\emph{cp}}} & Facebook & \tikz[baseline=-0.6ex] \fill[semiadequate] circle (0.6ex); & Overuse of content type `Other' with simple label `Account' (75\%). \\ 
 & Instagram & \tikz[baseline=-0.6ex] \fill[semiadequate] circle (0.6ex); & Overuse of content type `Other' with simple label `Account' (86\%). \\ 
 & LinkedIn & \tikz[baseline=-0.6ex] \fill[semiadequate] circle (0.6ex); & Heavy overuse of `Other' content type (100\%), but detailed. \\ 
 & Pinterest & \tikz[baseline=-0.6ex] \fill[semiadequate] circle (0.6ex); & Heavy overuse of uninformative `Other' content type with label 'Pin' (99\% of all SoRs). \\ 
 & Snapchat & \tikz[baseline=-0.6ex] \fill[semiadequate] circle (0.6ex); & Heavy overuse of less informative 'Other violation of T\&C' category (93\%). \\ 
 & TikTok & \tikz[baseline=-0.6ex] \fill[adequate] circle (0.5ex); & No substantial missing data. \\ 
 & X & \tikz[baseline=-0.6ex] \fill[inadequate] circle (0.75ex); & All records on SoRs for non-illegal content are missing. \\ 
 & YouTube & \tikz[baseline=-0.6ex] \fill[adequate] circle (0.5ex); & No substantial missing data. \\ 
\midrule\addlinespace[2.5pt]
\multirow{8}{*}{\shortstack[c]{\texttt{TDB}\\\emph{cn}}} & Facebook & \tikz[baseline=-0.6ex] \fill[semiadequate] circle (0.6ex); & No SoR processed by fully automated means despite declaring heavy use of full automation in TR. \\ 
 & Instagram & \tikz[baseline=-0.6ex] \fill[semiadequate] circle (0.6ex); & No SoR processed by fully automated means despite declaring heavy use of full automation in TR. \\ 
 & LinkedIn & \tikz[baseline=-0.6ex] \fill[semiadequate] circle (0.6ex); & Remarkably fewer own-initiative T\&C actions and automated share compared to TR. Inconsisten labeling of `Other' content type. \\ 
 & Pinterest & \tikz[baseline=-0.6ex] \fill[semiadequate] circle (0.6ex); & Markedly fewer Art. 16 actions and remarkably lower automated share of own-initiative actions compared to TRs. Highest SoR rate but no full automation. \\ 
 & Snapchat & \tikz[baseline=-0.6ex] \fill[semiadequate] circle (0.6ex); & Remarkably fewer Art. 16 actions compared to and remarkably less own-initiative T\&C actions compared to TRs. \\ 
 & TikTok & \tikz[baseline=-0.6ex] \fill[semiadequate] circle (0.6ex); & Markedly fewer Art. 16 actions compared to TRs. \\ 
 & X & \tikz[baseline=-0.6ex] \fill[inadequate] circle (0.75ex); & Supiscious almost perfect zero-day delay in SoRs with practically no automated means. Only illegal content, unlike TR. \\ 
 & YouTube & \tikz[baseline=-0.6ex] \fill[semiadequate] circle (0.6ex); & Remarkably fewer notice actions (especially by trusted flaggers) and own-initiative T\&C actions, as well as automation, compared to TR.\\ 
\midrule\addlinespace[2.5pt]
\multirow{8}{*}{\shortstack[c]{\texttt{TDB}\\\emph{tm}}} & Facebook & \tikz[baseline=-0.6ex] \fill[semiadequate] circle (0.6ex); & 2 days w/o submissions (1\% of 2025H2); late workload spread over circa 1 day. Marignal median delay (0.76 days). \\ 
 & Instagram & \tikz[baseline=-0.6ex] \fill[semiadequate] circle (0.6ex); & 2 days w/o submissions (1\% of 2025H2); late workload spread over circa 1 day. Marignal median delay (0.84 days). \\ 
 & LinkedIn & \tikz[baseline=-0.6ex] \fill[adequate] circle (0.5ex); & Submitted every day. Minimal median delay (almost 0.0 days). \\ 
 & Pinterest & \tikz[baseline=-0.6ex] \fill[semiadequate] circle (0.6ex); & 16 days w/o submissions. (8.7\% of 2025H2); late workload spread over circa 9 days. Marginal median delay (almost 0.0 days) \\ 
 & Snapchat & \tikz[baseline=-0.6ex] \fill[adequate] circle (0.5ex); & Submitted every day. Minimal median delay (almost 0.0 days). \\ 
 & TikTok & \tikz[baseline=-0.6ex] \fill[inadequate] circle (0.75ex); & 36 days w/o submissions (19.5\% of 2025H2); late workload spread over circa 30 days. Marked median delay (17.6 days). \\ 
 & X & \tikz[baseline=-0.6ex] \fill[inadequate] circle (0.75ex); & 52 days w/o submissions (28.2\% of 2025H2); late workload spread over circa 3 months. Remarkable median delay (30.9 days). \\ 
 & YouTube & \tikz[baseline=-0.6ex] \fill[adequate] circle (0.5ex); & Submitted every day. Minimal median delay (1.0 days). \\ 
\bottomrule
\end{tabular*}
\caption{Empirical evidence underlying the structured comparative assessment reported in Table~\ref{tab:platform_evaluation}. For each platform, transparency mechanism, and applicable reporting dimension, the table reports the assigned ordinal adequacy level together with the main empirical observations supporting the assignment. Dimensions are timeliness (\emph{tm}), consistency (\emph{cn}), completeness (\emph{cp}), and formatting (\emph{fm}). Colored dots represent  assessment: \emph{inadequate}~(\tikz[baseline=-0.6ex] \fill[inadequate] circle (0.7ex);), \emph{semi-adequate}~(\tikz[baseline=-0.6ex] \fill[semiadequate] circle (0.5ex);), and \emph{adequate}~(\tikz[baseline=-0.6ex] \fill[adequate] circle (0.4ex);).}
\label{tab:platform_evaluation_commented}
\end{table}